Ellen D. Williams
Department of Physics,
University of Maryland,
College Park, Maryland 20742-4111
Tel: 301-405-6156
Fax: 301-314-9465
e-mail: edw@physics.umd.edu1

# Nanoscale Equilibrium Crystal Shapes

*M. Degawa, F. Szalma and E.D. Williams*

*Department of Physics,
Materials Research Science and Engineering Center,
University of Maryland,
College Park, Maryland 20742-4111*

**Abstract**

The finite size and interface effects on equilibrium crystal shape (ECS) have been investigated for the case of a surface free energy density including step stiffness and inverse-square step-step interactions.  Explicitly including the curvature of a crystallite leads to an extra boundary condition in the solution of the crystal shape, yielding a family of crystal shapes, governed by a shape parameter $c$.  The total crystallite free energy, including interface energy, is minimized for $c=0$, yielding in all cases the traditional PT shape ($z \sim x^{3/2}$). Solutions of the crystal shape for $c \neq 0$ are presented and discussed in the context of meta-stable states due to the energy barrier for nucleation. Explicit scaled relationships for the ECS and meta-stable states in terms of the measurable step parameters and the interfacial energy are presented.

## 1. Introduction

Technological demands for the fabrication of nano-structures and quantum dots [1-3] provides renewed motivation for understanding the atomistic properties that control the morphology of crystal shapes. With decreasing structure size, the issues of finite size and shape effects become non-negligible [4,5] including issues of stability against decay and structural rearrangement [6-8]. Small structures are also increasing sensitive to external perturbations, such as the stress caused by the substrate interface [9-13]. In this work we will address the issues of nanoscale equilibrium crystal shape for the surface free energy of the continuum step model [14,15], which allows experimentally determined thermodynamic parameters for step free energies to be connected rigorously to structural predictions.

The theory of equilibrium crystal shapes (ECS) has been extensively studied in the last half century [16-25]. The formation of facets below the roughening temperature, and the Pakrovsky-Talapov predictions [26] for the edge shape ($z(x)$) of crystals have been demonstrated in a number of clean systems [27-34]. In addition to studies on 3D crystals, there has been substantial work on the 2D ECS of islands [15,35-39] to obtain the edge free energy, equilibrium edge fluctuations and decay kinetics.

While 2D studies have intrinsically addressed the effects of finite size, most theoretical 3D studies have been performed in the limit of large crystal size, where curvature effects can be neglected. However, indications of size effects have been presented in theoretical work on nucleation barriers in crystal evolution [25,40,41] and shape [2,42]. Furthermore, curvature effects are clearly important in the stability or evolution on 3D crystallites [43-46]. Moreover, interesting new fluctuation phenomena are possible in small crystallites, as shown theoretically by Ferrari et al.[47].

Here we explicitly consider the case of a "free standing" truncated crystal, as in Fig. 1 a) or supported truncated crystallite, Fig. 1 b). We address the nature of the equilibrium shape for these models, as a first step in describing the evolution of crystallites under external perturbation [48]. The effects of finite curvature and interfacial effects are presented quantitatively in terms of experimentally measurable parameters.

## 2. Background

The equilibrium crystal shape arises as the result of the minimization of the orientation-dependent surface free energy $\gamma(\hat{n})$ with the constraint of constant volume. This calculation is performed via the Wulff construction [17,49], which mathematically is a Legendre transform from the surface tension to the ECS. Vice versa the ECS provides a direct measurement of the surface tension. Interface effects in ECS have been considered for T=0 crystal shapes by Muller and Kern [9]. They have considered a crystal shape truncated at a bottom surface (with surface free energy $\gamma_A$ equal to that of the top facet) and brought into contact with a substrate surface of free energy $\gamma_B$. They use the Dupré relationship $\gamma_{AB} = \gamma_A + \gamma_B - E_A$ for the interface free energy $\gamma_{AB}$, where $E_A$ is the adhesion energy. This yields a change in the substrate free energy per unit area due to formation of the interface of $\gamma_{AB} - \gamma_B = \gamma_A - E_A$. They show that when the substrate free energy change is zero, e.g. $\gamma_A - E_A = 0$, the crystal truncation occurs at the Wulff point, e.g. the configuration shown in Fig. 1 a). For larger values of the binding energy, the truncation occurs as shown in Fig. 1 b), with the crystallite height $z_h$ proportional to $2\gamma_A - E_A$.

Most studies of the rounded edges of crystals have used the limit of infinite size, in which curvature effects are neglected, and thus yield an edge profile $z(x)$ independent of the third dimension [25]. For finite size crystallites, curvature effects of the steps become non-negligible, thus the full profile $z(x,y)$ must be considered. The full 3D Wulff construction can be presented in rectangular coordinates [50], however a more efficient way to take curvature effects into account is to work in cylindrical coordinates yielding $z(r,\theta)$. The 3D Wulff construction in cylindrical coordinates is [37]

$$\gamma(\hat{n}) = \left(\frac{\lambda}{2}\right) \frac{z - r\left(\frac{\partial z}{\partial r}\right)}{\left[1 + \left(\frac{\partial z}{\partial r}\right)^2 + \frac{1}{r^2}\left(\frac{\partial z}{\partial \theta}\right)^2\right]^{1/2}}, \qquad (1)$$

where $\lambda$ is the Legendre multiplier, which turns out to be the *excess* chemical potential due to the surfaces. For the surface tension $\gamma(\varphi)$, or the surface free energy density $f(\varphi)$, let us consider the Pokrovsky-Talapov type surface free energy density [26], which is a good approximation for surfaces which make a low angle $\varphi$ with respect to a neighboring low-index facet orientation:

$$\frac{\gamma(\hat{n})}{\cos\varphi} = f(\varphi) = \gamma_0(T) + \frac{\beta(\theta,T)}{h}\tan\varphi + g(\theta,T)\tan^3\varphi. \qquad (2)$$

Here $h$ is the step height, $\beta$ and $g$ are the thermodynamic step free energy and step-step interaction coefficient, respectively, $\gamma_0$ is the surface tension of the low-index terrace and $\varphi$ is the angle of the surface relative to that terrace, thus $\tan\varphi = \partial z/\partial r$ corresponds to the step density. With an isotropic step free energy (no $\theta$ dependence, $\beta=\tilde{\beta}$), this Wulff construction gives the same solution as the Pokrovsky-Talapov equilibrium crystal shape (PT-ECS) for an infinite volume in 1D. The only difference is the factor of 2 in the denominator of the Legendre multiplier in eq.(1) due to the dimensionality [37].

In addressing the evolution of crystal shapes, it is useful to address the ECS by defining a local excess chemical potential that must be the same everywhere on the crystal surface. The local surface chemical potential is obtained by calculating the change in total free energy when there is a small local deformation of the surface, e.g. one must take the functional derivative $\delta F / \delta N$. Here, we evaluate this derivative using the continuum model, where the surface is approximated as a 2D continuous interface. This is usually used for a crystal shape above the roughening temperature, however it is still valid for the rounded regions on a crystal with facets that are below their roughening temperature.

For a single-component incompressible ($V=vN$) crystallite, the Helmholtz free energy $F(T,N)$ is the most convenient ensemble to consider. For an arbitrary surface free energy density $f$, the total free energy of the crystallite in cylindrical coordinates is

$$F = \iint f(z_r, \frac{z_\theta}{r}) r dr d\theta + \frac{\mu_B}{v} \iint z(r,\theta) r dr d\theta + (\gamma_0 - E_A) \iint r dr d\theta, \qquad (3)$$

where $z_r = \partial z/\partial r$ is the local slope of the surface, $v$ the atomic volume and $\mu_B$ is the fixed chemical potential of the bulk. The first term is the surface free energy, the second term is a constant bulk free energy and the third term is the free energy of the truncated bottom surface including the adhesion energy at the interface if there is any. The equilibrium configuration minimizes the total free energy at fixed N, and yields a uniform value of the excess chemical potential

$$\frac{\delta F}{\delta N} = -\mu_S(r,\theta) + \mu_B = \mu(V). \qquad (4)$$

From eq. (3) and (4), the local excess chemical potential can be defined as

$$\mu_S(r,\theta) = v \left[ \frac{1}{r} \frac{\partial f}{\partial z_r} + \frac{d}{dr}\left(\frac{\partial f}{\partial z_r}\right) + \frac{d}{d\theta}\left(\frac{\partial f}{\partial z_\theta}\right) \right]. \qquad (5)$$

Here let us use eq. (2) for the surface free energy density. We also neglect any variation along the step edge direction ($\beta = \check{\beta}$) by dropping the third term in eq. (5), e.g. assuming cylindrical symmetry as shown in Fig. 1. This gives the dependence of $\mu_S(r)$ for an arbitrary crystallite shape:

$$\mu_S(r) = \frac{\Omega\beta(T)}{r} + \frac{3vg(T)}{r}\left(\frac{dz(r)}{dr}\right)^2 + 6vg(T)\left(\frac{dz(r)}{dr}\frac{d^2z(r)}{dr^2}\right). \tag{6}$$

The first term in eq. (6) is the 2D Gibbs-Thomson term originating from the curvature of the edge of the top facet and the corresponding curvature of the crystal edge. The second term is the change in step interaction energy due to the change of the circumference of the interacting step edge when a single atom is removed from the layer edge. The last term is a change in step interaction energy, which is obtained also in the limit of zero curvature (straight steps). The first two terms vanish as r→ ∞, as must be the case.

### 3. Results

The steady state crystallite shape (e.g. a shape where $\mu_S(V)$ is everywhere the same) can be obtained from eq. (6), where the excess chemical potential $\mu_S(V) = \mu_B - \mu(V)$ will be determined by the volume of the crystallite. Setting $y(r) = (dz(r)/dr)^2$ and performing the integration once gives

$$\frac{dz(r)}{dr} = -\sqrt{\frac{\mu_S(V)r - 2\beta\Omega}{6gv} + \frac{c}{r}}, \tag{7}$$

where $c$ is a constant of integration that appears due to including the third dimension (curvature) in the calculation. The minus sign is taken to conform with the coordinate system shown in Fig. 1. It is possible to find the relationship between the facet radius $\rho_c$ (see Fig. 1) and the excess chemical potential $\mu_S(V)$ simply by setting $dz(r)/dr\,|_{\rho c}=0$. This yields:

$$\rho_c = \frac{\beta\Omega + \sqrt{\beta^2\Omega^2 - 6cgv\mu_S(V)}}{\mu_S(V)} \quad \text{or} \quad \mu_S(V) = \frac{2(\beta\rho_c\Omega - 3cgv)}{\rho_c^2} \tag{8}$$

### 3.1. Pakrovsky-Talapov Shape

At this point, it is possible to reproduce the known results for the PT-ECS, which is a specific solution of eq.(7). In order to satisfy one component of the Wulff relation, $\rho_0=2\beta\Omega/\mu_S(V)$ the constant of integration $c$ must equal 0. Integrating eq. (7) with $c=0$ yields:

$$z(r) = z_0 - \sqrt{\frac{2\mu_s(V)}{27gv}}\left(r - \frac{2\Omega\beta}{\mu_S(V)}\right)^{3/2}, \tag{9}$$

where $z_0$ is taken to be the distance between the facet and the center of the crystallite (the Wulff point) to satisfy the other component of the Wulff relation $z_0=2\gamma_0 v/\mu_S(V)$. So

$$z(r) = z_0 - \frac{2}{3}\sqrt{\frac{\beta}{3gh\rho_0}}(r - \rho_0)^{3/2} \qquad \frac{\gamma_0}{z_0} = \frac{\beta}{h\rho_0}\left(=\frac{\mu_S(V)}{2v}\right). \tag{10}$$

Thus the solution with the boundary conditions ($z_h=z_0$, $c=0$) that gives the Wulff relation, also gives the Pakrovsky-Talapov ECS[23,25].

By defining the crystallite so that its truncated interface passes through the Wulff point as in Fig. 1 a) ($z_h=z_0$), we are using the free standing boundary condition where the top (and bottom) facet surface tension is equal to the adhesion energy [9]. For this case, we can calculate the radius of the interface area $R_b$ and also the contact angle to the interface plane $\varphi_b$ analytically, from $z(R_b)=0$ and $\tan\varphi_b=dz(r)/dr\,|_{Rb}$:

$$R_b = \frac{2\beta\Omega + 3v(2g\gamma_0^2)^{1/3}}{\mu_s(V)} \quad \text{and} \quad \tan\varphi_b = \left(\frac{\gamma_0}{2g}\right)^{1/3} \tag{11}$$

An explicit expression for the excess chemical potential can now be precisely determined by the volume constraint:

$$\pi\rho_0^2 z_0 + \int_0^{2\pi}\int_{\rho_0}^{R_b} z(r)r\,dr\,d\theta = V = const.$$

yielding:

$$\mu_s(V) = \Omega\left(\frac{2\pi h\gamma_0(140\beta^2 + 252\beta h(2g\gamma_0^2)^{1/3} + 135h^2(2g\gamma_0^2)^{2/3})}{35V}\right)^{1/3}. \tag{12}$$

A similar set of expressions can be obtained for a supported truncated crystallite as shown in Fig. 1 b). This is the case where the adhesion energy is not equal to the facet surface free energy $E_A \neq \gamma_0$, and thus the Wulff point is *not* located in the truncation plane of the crystallite, $z_h \neq z_0$. Following Muller's arguments [9], the second Wulff relationship is replaced by the Wulff-Kaishew relation

$$\frac{\mu_S(E_A,V)}{2v} = \frac{2\gamma_0 - E_A}{z_h} = \frac{\beta}{h\rho_0}. \qquad (13)$$

This fixes the contact angle to the substrate $\varphi_b$ and also the excess chemical potential is obtained as a function of adhesion energy and volume.

$$\tan\varphi_b = \left.\frac{dz}{dr}\right|_{R_b(z_0-z_h)} = \left(\frac{2\gamma_0 - E_A}{2g}\right)^{1/3} \;\rightarrow\; z_h(E_A) = \frac{2gh\rho_0 \tan^3\varphi_b}{\beta},$$

$$\mu_S(E_A,V) = 2\Omega \tan\varphi_b \left(\frac{2\pi gh(35\beta^2 + 126\beta gh \tan^2\varphi_b + 135g^2h^2\tan^4\varphi_b)}{35V}\right)^{1/3}. \qquad (14)$$

The phase space of parameters governing the ECS can be reduced by noting a common dependence on $2\gamma_0$-$E_A$, as shown in Appendix A, eqs. (A-1) and (A-2). This reduces eqs. (12) and (14) to a common form

$$\tilde{\mu}_S = \left(\frac{2\pi(140 + 252\sqrt[3]{2}\tilde{E}_A + 135\sqrt[3]{2^2}\tilde{E}_A^2)}{35\tilde{V}}\right)^{1/3} \;\text{ where }\; \tilde{E}_A = \frac{(g(2\gamma_0 - E_A)^2)^{1/3}}{\beta/h}. \qquad (15)$$

Fig. 2 a) shows the dimensionless excess chemical potential $\tilde{\mu}_S$ as a function of the dimensionless volume $\tilde{V}=V/h\Omega$ with different physically reasonable values of $\tilde{E}_A$. The range of $\tilde{E}_A$ from 1.0~2.0 was chosen to illustrate the estimated variations expected for Pb(111) for a temperature range from approximately room temperature to the melting temperature [32,33]. Fig. 2 b) gives the dimensionless facet radius $\tilde{\rho}_0$ (or height $\tilde{z}_h$), contact slope $\tan \tilde{\varphi}_b$, and interface radius $\tilde{R}_b$ of the crystallite as a function of the ratio $\tilde{E}_A$ at constant volume $\tilde{V}$. The scaled facet radius and crystal height reduces to the same function, however the physical (unscaled) facet radius and height vary with temperature (or adhesion energy) quite differently. Specially, with all other parameters held constant the height decreases with increasing adhesion energy, where the facet radius increases.

### 3.2. Non- Pakrovsky-Talapov Shapes

The results above have all been obtained with the shape parameter $c$ of eq. (7) equal to zero. In the case of infinite (zero curvature) crystallites, this is the only solution, and it yields the ECS. However, in this case, there is an entire family of solutions, only one of which corresponds to the ECS. (E.g. it is possible for the whole surface to have a constant surface chemical potential and yet not be the minimum total free energy configuration that corresponds to the ECS.) Before discussing the physical meaning of the solutions with $c \neq 0$, we note from eq. (8) that there is a maximum possible value of $c$. This is simply obtained by setting the term in the root equal to 0.

$$c_{max} = \frac{\beta^2 \Omega}{6gh\mu_S(c_{max},V)} = \frac{\beta \rho_{c\,max}}{6gh} \qquad (16)$$

At this maximum value of $c$, $\mu_S(c_{max},V)=\beta\Omega/\rho_{cmax}$. With this value of $c_{max}$, eq. (7) can be easily solved and gives a crystallite shape of

$$z(r) = z_h - \frac{2\beta\Omega}{3\mu_S(c_{max},V)}\sqrt{\frac{2\beta}{3gh}} - \frac{1}{3}\sqrt{\frac{2\mu_S(c_{max},V)r}{3gv}}\left(r - \frac{3\beta\Omega}{\mu_S(c_{max},V)}\right)$$

$$= z_h - \frac{2\rho_{cmax}}{3}\sqrt{\frac{2\beta}{3gh}} - \frac{1}{3}\sqrt{\frac{2\beta r}{3gh\rho_{cmax}}}(r - 3\rho_{cmax}) \qquad (r>\rho_{cmax})$$

$$\cong z_h - \frac{1}{3}\sqrt{\frac{2\beta}{3gh}}\left(\frac{3(r-\rho_{cmax})^2}{4\rho_{cmax}} - \frac{(r-\rho_{cmax})^3}{12\rho_{cmax}^2} + O^4\right) \qquad (17)$$

As previously noted by Uwaha [23], $c=c_{max}$ at the vicinal surface leading to the facet has a parabolic shape $z \sim x^2$ in contrast to the PT-ECS where it has an $x^{3/2}$ dependence. At other values of the shape parameter, the crystallite shape must be determined numerically. Taking the free standing boundary condition ($E_A=\gamma_0$), Fig. 3 shows size scaled cross sections $\tilde{z}(\tilde{r})$ of the crystallites with the same volume but different ratios of the scaled shape parameter (scaling given in the Appendix eq. (A-3) and (A-4) yields $\tilde{c}_{max}=1/6$). The shape changes from the parabolic shape for $\tilde{c}$ close to $\tilde{c}_{max}$ to the 3/2 exponent as the shape parameter decreases to zero. For arbitrary values for $c$ and $E_A$, the crystal shapes can be obtained numerically. The results can be compiled using scaling so that they can be used for any chosen set of parameters, as shown in Fig. 4. Fig. 4 a) gives scaled (eq. (A-3), (A-5) and (A-6)) crystallite height $\tilde{z}_h$, b) interface radius $\tilde{R}_b$, c) contact slope $\tan\tilde{\varphi}_b$ and d) crystallite volume $\tilde{V}$ as a function of the shape parameter $\tilde{c}$ up to 1/6 with constant volume. The scaling works perfectly in all cases when $\tilde{c}=0$. However, such parameters as $\tilde{R}_b$ and $\tilde{V}$ do not scale perfectly when $\tilde{c}\neq 0$. The ratios of the crystallite height, interface radius and volume to the facet radius are increasing functions of the shape parameter. The slope on the other hand decreases with shape parameter independent of the volume. Given the expression for the shape of the crystallite, it is possible once again to calculate the constraint of the volume to obtain the excess chemical potential in terms of the other physical parameters. The results for $\tilde{\mu}_S$ is shown also as a function of the scaled shape parameter in Fig. 4 e). At constant volume, it is a decreasing function of the shape parameter.

### 3.3. Equilibrium Shape and Metastable Shape

Given the solutions above, we are now in a position to determine which value of $c$ yields the minimum free energy for any value of the adhesion energy and thus the ECS. We begin with the crystallite free energy eq. (3)

$$F_S(c,V) = 2\pi \int_0^{Rb} f(r,c,\mu_S(c,V))r dr + \pi(\gamma_0 - E_A)R_b^2 + const.$$

$$= \pi \gamma_0 \rho_c^2(c,\mu_S(c,V))$$

$$+ 2\pi \int_{\rho c}^{Rb} \left( \gamma_0 + \frac{\beta}{h}\sqrt{\frac{\mu_S(c,V)r - 2\beta\Omega}{6gv} + \frac{c}{r}} + g\left(\frac{\mu_S(c,V)r - 2\beta\Omega}{6gv} + \frac{c}{r}\right)^{3/2} \right) r dr$$

$$+ \pi(\gamma_0 - E_A)R_b^2 + const. \qquad (18)$$

Fig. 5 a) gives the surface free energy (the first two terms in eq. (18)) as a function of $c$ up to $c_{max}$ with constant $V$ for the free standing boundary condition ($E_A=\gamma_0$). The thermodynamic parameters where calculated using parameters for Pb at 27 ˚C, $\gamma_0=1.7$ eV/nm$^2$[51], $\beta=0.34$ eV/nm and $g=0.65$ eV/nm$^2$ [32,33]. The value $c=0$ clearly gives the absolute minimum of the surface free energy and thus also of the total crystallite free energy, thus yielding the ECS. The solutions with $c \neq 0$ are not absolute minima in the free energy, and thus since they have a constant surface chemical potential, they may represent meta-stable physical configurations.

Finally, let us consider the physically significant case of a heterogeneous crystallite-substrate system ($E_A \neq \gamma_0$) with $c \neq 0$. Combining eqs. (8) and (13), the boundary condition is now

$$\frac{\mu_S(c, E_A, V)}{2v} = \frac{\beta \rho_c \Omega - 3cgv}{v \rho_c^2} = \frac{2\gamma_0 - E_A}{z_h}$$

As a result, the excess chemical potential is now a function of the shape parameter, the adhesion energy and volume. In this case the surface and total free energies do not have the same $c$ dependence. Here we consider first the surface free energy then the total crystallite free energy (e.g. including the interface terms). Filled triangles and squares in Fig. 5 b) and c) give the surface free energy as a function of the shape parameter for the same volume crystallites but for different adhesion energies $E_A=0.5\gamma_0$ and $E_A=1.5\gamma_0$, respectively. Clearly, the minimum of the surface free energy shifts from zero with different shape parameters. For stronger adhesion, $E_A=1.5\gamma_0$, the minimum occurs at larger $c$, e.g. closer to the critical state, and for weaker adhesion, $E_A=0.5\gamma_0$, the shift is in the c<0 direction. The larger the adhesion energy, the greater the tendency toward a larger interface area, resulting in a flatter crystallite when the volume is fixed. This flattening gives smaller contact slopes at the interface (see eq.(14) for $c=0$), which results in a shift of the minimum of the surface free energy in the $c_{max}$ direction. Similarly, when the adhesion energy is small, the interface area is small and the contact slope is large, which then shifts the minimum of the surface free energy in the $c<0$ direction. However, when adding on the interface energy to obtain the total crystallite free energy the minimum of the crystallite free energy is shifted back to $c=0$ independent of the adhesion energy. This is shown in Fig. 5 b) and c) as open triangles and squares. Thus, when the total crystallite excess free energy is considered, the PT-ECS always stands independent of the adhesion energy. Notice that the crystallite free energy is the smallest when the adhesion energy is largest, as expected, however this is not the case when considering only the surface free energy. This indicates that for calculating the properties of nano-size crystallites you must minimize the total crystallite excess free energy including the interface and not only the surface free energy. The specific free energy curves of Fig. 5 b) and c) can be generalized by using the scaled (eqs. (A-3) and (A-6)) parameters as shown in Fig. 5 d). All the curves collapse to one point at $c=0$. This result shows that we can relatively calculate the ECS using eqs. (A-1) and (A-2). Thus predicting variations in shape, due to e.g. changes in adhesion energy, can be done simply and relatively.

If we are interested in the possible metastable state represented by $c\neq0$, then the use of the scaling forms of eqs. (A-3) is necessary, along with the function relationships of eqs. (A-4) and (A-5), which are represented in Fig. 3, 4 and 5. Given, for instance experimental information as crystal height and facet radius, graphical prediction of Fig. 4, (or the values in Table 1) all the other scaled parameters can be determined. Analytical solutions (eqs. (A-5)) yield excellent scaling results for crystal height, contact slope and excess chemical potential (Fig. 4 a), c) and e)). The interface radius, the volume and total free energy have not yielded robust scaling relationships (Fig. 4 b), d) and 5 d)) and thus must be calculated individually if quantitative values are desired.

### 3.4. Application to Experiment

The consequence of the potential meta-stable states can be evaluated by comparison of the predicted crystal shape profiles with experimentally measured results. Fig. 6 shows a log-log plot of a 3 parameter fit ($z_h$, $\rho$, $\beta/g$) of eq. (10) (the ECS) and eq.(17) (the metastable state at $c_{max}$) to a cross section of a defect-free Pb crystallite measured with an STM at ~ 27 °C [45]. The solid line is the fit to a PT-shape and the dashed line to the critical state shape. Both shapes visually fit the crystal shape well. However, the ratios of the interaction coefficient to the step free energy $g/\beta$ obtained from the fits are very different, $g/\beta$=17.2 nm$^{-1}$ for the PT-ECS and $g/\beta$=3.6 nm$^{-1}$ for the critical state fit. This presents an interesting comparison with experiments by Nowicki et al. [33], where they have used fits to a PT-ECS and obtained the ratio of $g/\beta$=4.96 nm$^{-1}$ for an equilibrated crystallite and $g/\beta$=13.11 nm$^{-1}$ for a non-equilibrated crystallite at similar temperatures. (In their experiment they characterize crystallites that have screw dislocations on the facet as equilibrated as these structures can rearrange without the need to overcome a nucleation barrier [40,41,45].) Since the crystallite fit in Fig. 6 did not have such a dislocation, we can speculate that the PT-ECS fit in this case and in Nowicki's non-equilibrated case both yield a spuriously high value of g, as previously predicted by Thürmer [45], because the crystallites are trapped in meta-stable ($c \neq 0$) states. The fit to the parabolic state seems to correct properly for the possible meta-stable structure as it yields a value of $g/\beta$ consistent with these measured for crystallites that have relaxed via motion of screw dislocation.

## 4. Discussion

The use of the physically-based functional form of eq. (3) for the orientation dependence of the surface free energy yields a ready formalism for evaluating the behavior of nanoscale crystallites supported on substrates of varying interaction strength. The use of the scaling formulas developed in the Appendix makes the results easy to extend to variable material systems. As already noted by Muller and Kern [9], increasing binding energy yields more severely truncated crystal shapes. Here we explicitly show also how the shape at the facet edge will respond to such substrate interactions: specifically the shift in the surface free energy minimum (shown in Fig. 5 b) and c)) due to the presence of the interface. This would seem to indicate the break down of the Wulff theorem, however, when the interface energy is added to obtain the total crystallite free energy, the minimum of the total free energy is shifted back to $c=0$, the PT-ECS. This is physically reasonable, for large adhesion energy a large interface area is desired leading to a shift in the negative direction and for small vice versa. Thus we obtain an interesting result that the Wulff theorem gives the minimum of the total free energy even for finite crystallites and also with interfaces. The interesting effect of substrate interactions on the edge shapes are a direct consequence of the finite size effect, e.g. the new behavior that results from explicitly considering the shape of the crystallite that becomes important at small volumes.

Explicitly considering the effects of the curvature yields a family of crystal shapes that all have constant surface chemical potentials, only one of which represents an absolute minimum in the total surface free energy. The others, because they will have no strong driving force for rearrangement by mass transport, are likely to represent meta-stable states. This conclusion gives interesting results, such as on the vapor-solid coexistence curve: though the vapor (and bulk) chemical potentials may be fixed, stable crystallites with different numbers of atoms (volume) will still be possible due to the difference in shape parameter, although only one of the shapes (volume) is the ECS.

The possibility of meta-stable crystal shapes, trapped by barriers to rearrangement has been proposed by Rohrer and Mullins [40,41]. The barriers to shape evolution $\Delta E$ can be estimated by the method they proposed [25,40,41] considering only formation of steps

$$\Delta E(r) = 2\pi\beta(r - \rho_c) + \frac{\pi\mu_S(V)(\rho_c^2 - r^2)}{\Omega} \tag{18}$$

The critical radius is $r^* = \beta\Omega/\mu_b$ and with eq.(8) the barrier height is

$$\Delta E(r_C) = \frac{\pi(\beta\rho_c\Omega - 6cgv)^2}{2\Omega(\beta\rho_c\Omega - 3cgv)} \tag{19}$$

Fig. 7 shows the scaled barrier height $\Delta \tilde{E}$ as a function of $\tilde{c}$ (eq. (A-3) and (A-5))for different adhesion energies. It can be seen that the barrier disappears at $\tilde{c}_{max}$ =1/6, in agreement with Uwaha's identification of this point as a "critical state"[23]. The barrier can be understood in more detail by working in the continuum step model where the concept of steps is included. This will be presented elsewhere [48].

**Acknowledgments:** This work has been supported by the UMD-NSF MRSEC under grant DMR 00-80008. We gratefully acknowledge useful discussions with Prof. M. Uwaha, Prof. J.D. Weeks and Prof. A. Pimpinelli.

**Appendix A**

For the PT-ECS ($c=0$) all calculations can be done analytically, and size scaling can be done using the volume of the crystallite. We define dimensionless quantities of the adhesion energy (temperature), volume, excess chemical potential, facet radius, crystallite height, interface radius and contact angle as follows

$$\widetilde{E}_A = \frac{(g(2\gamma_0 - E_A)^2)^{1/3}}{\beta/h}, \quad \widetilde{V} = \frac{V}{h\Omega}, \quad \widetilde{\mu}_S = \frac{\mu_S}{(\Omega^2(2\gamma_0 - E_A)\beta^2)^{1/3}},$$

$$\widetilde{\rho}_0 = \frac{\rho_0}{(\beta\Omega/(2\gamma_0 - E_A))^{1/3}}, \quad \widetilde{z}_h = \frac{z_h}{h((2\gamma_0 - E_A)^2\Omega/\beta^2)^{1/3}},$$

$$\widetilde{R}_b = \frac{R_b}{(\beta\Omega/(2\gamma_0 - E_A))^{1/3}}, \quad \tan\widetilde{\varphi}_b = \frac{\tan\varphi_b}{(\beta/gh)^{1/2}}. \tag{A-1}$$

This gives the following analytical relationships between dimensionless variables, which are shown in Fig. 2

$$\widetilde{\mu}_S = \left(\frac{2\pi(140 + 252\sqrt[3]{2}\widetilde{E}_A + 135\sqrt[3]{2^2}\widetilde{E}_A^2)}{35\widetilde{V}}\right)^{1/3},$$

$$\widetilde{\rho}_0 = \widetilde{z}_h = \left(\frac{140\widetilde{V}}{\pi(140 + 252\sqrt[3]{2}\widetilde{E}_A + 135\sqrt[3]{2^2}\widetilde{E}_A^2)}\right)^{1/3},$$

$$\widetilde{R}_b = (2 + 3\sqrt[3]{2}\widetilde{E}_A)\left(\frac{35\widetilde{V}}{2\pi(140 + 252\sqrt[3]{2}\widetilde{E}_A + 135\sqrt[3]{2^2}\widetilde{E}_A^2)}\right)^{1/3},$$

$$\tan\tilde{\varphi}_b = \frac{\sqrt{E_A}}{\sqrt[3]{2}}. \tag{A-2}$$

For $c \neq 0$ states size scaling is not done in terms of the volume but in terms of the characteristic facet radius at each state numerically calculated for constant volume. Here we define dimensionless values of position, shape parameter, cross section, interface radius, crystal height, contact slope, excess chemical potential, crystallite volume, crystallite free energy and energy barrier $\Delta E$ as

$$\tilde{r} = \frac{r}{\rho_c}, \quad \tilde{c} = \frac{cgh}{\rho_c \beta}, \quad \tilde{z}(\tilde{r}) = \frac{z(r) - z_h}{\rho_c (\beta/gh)^{1/2}}, \quad \tilde{R}_b = \frac{R_b}{\rho_c},$$

$$\tilde{z}_h = \frac{z_h \beta}{\rho_c h(2\gamma_0 - E_A)}, \quad \tan\tilde{\varphi}_b = \frac{\rho_c R_b gh \tan^2 \varphi_b}{\beta(R_b^2 - \rho_c^2)} - \frac{R_b}{3(R_b + \rho_c)}, \quad \tilde{\mu}_S = \frac{\rho_c \mu_S}{\beta \Omega},$$

$$\tilde{V} = \frac{V\beta}{\rho_c^3 h(2\gamma_0 - E_A)}, \quad \tilde{F} = \frac{F}{\rho_c^2 (2\gamma_0 - E_A)}, \quad \Delta\tilde{E} = \frac{\Delta E}{\rho_c \beta}. \tag{A-3}$$

Notice the consistency with eq. (A-1) at $c=0$. Eqs. (A-3) yield specific expressions for the crystal shape for arbitrary values of $c$:

$$\tilde{z}(\tilde{r}) = -\int \sqrt{\frac{\tilde{r}-1}{3} - \frac{\tilde{c}(\tilde{r}^2-1)}{\tilde{r}}} d\tilde{r}, \tag{A-4}$$

$$\tilde{z}_h = \frac{1}{1-3\tilde{c}}, \quad \tilde{\mu}_S = 2(1-3\tilde{c}), \quad \tan\tilde{\varphi}_b = -\tilde{c}, \quad \Delta\tilde{E} = \frac{\pi(1-6\tilde{c})^2}{2(1-3\tilde{c})}, \tag{A-5}$$

$$\tilde{V} = \pi\left\{\frac{\tilde{R}_b^2}{1-3\tilde{c}} - 2\tilde{E}_A^{-3/2}\int_1^{\tilde{R}_b}\left(\int_1^{\tilde{r}}\sqrt{\frac{\tilde{r}'-1}{3} - \frac{\tilde{c}(\tilde{r}'^2-1)}{\tilde{r}'}}d\tilde{r}'\right)\tilde{r}d\tilde{r}\right\},$$

$$\tilde{F} = \pi\left\{\tilde{R}_b^2 + 2\tilde{E}_A^{-3/2}\int_1^{\tilde{R}_b}\left[\sqrt{\frac{\tilde{r}-1}{3} - \frac{\tilde{c}(\tilde{r}^2-1)}{\tilde{r}}} + \left(\frac{\tilde{r}-1}{3} - \frac{\tilde{c}(\tilde{r}^2-1)}{\tilde{r}}\right)^{3/2}\right]\tilde{r}d\tilde{r}\right\}.$$

eq. (A-4) is shown in Fig. 3 and eq. (A-5) are shown in Fig. 4 and 7. Notice that eq. (A-5) is already independent of adhesion energy (temperature), however, quantities such as the interface radius, crystallite volume and crystallite free energy are not. Also because analytical results are not available here we have used the scaling results obtained for $c=0$ (eq. (A-2)) and applied them to other shape parameters.

$$\tilde{R}_b = \frac{R_b - \rho_c}{\rho_c\tilde{E}_A},$$

$$\tilde{V} = \frac{V\beta}{\rho_c^3 h(2\gamma_0 - E_A)\tilde{E}_A^2} - \frac{\pi(5 + 9\sqrt[3]{2}\tilde{E}_A)}{5\tilde{E}_A^2}$$

$$\tilde{F} = \frac{F}{\rho_c^2(2\gamma_0 - E_A)\tilde{E}_A^2} - \frac{3\pi(5 + 9\sqrt[3]{2}\tilde{E}_A)}{5\tilde{E}_A^2}, \tag{A-6}$$

eq. (A-5) are shown in Fig. 4 and 5. As is discussed in the manuscript, this is not a perfect scaling and some temperature dependence remains.

# References

[1] Z. Gai, B. Wu, J.P. Pierce, G.A. Farnan, D.J. Shu, M. Wang, Z.Y. Zhang, J. Shen, Physical Review Letters 89 (2002) 235502.
[2] A.A. Golovin, S.H. Davis, P.W. Voorhees, Physical Review E 68 (2003) 056203.
[3] I. Lyubinetsky, S. Thevuthasan, D.E. McCready, D.R. Baer, Journal of Applied Physics 94 (2003) 7926.
[4] Z. Hens, D. Vanmaekelbergh, E.J.A.J. Stoffels, H. van Kempen, Physical Review Letters 88 (2002) 236803.
[5] S.J. Prado, C. Trallero-Giner, A.M. Alcalde, V. Lopez-Richard, G.E. Marques, Physical Review B 69 (2004) 201310.
[6] M. Uwaha, K. Watanabe, Journal of the Physical Society of Japan 69 (2000) 497.
[7] N. Israeli, D. Kandel, Physical Review B 60 (1999) 5946.
[8] A. Ichimiya, K. Hayashi, E.D. Williams, T.L. Einstein, M. Uwaha, K. Watanabe, Physical Review Letters 84 (2000) 3662.
[9] P. Muller, R. Kern, Surface Science 457 (2000) 229.
[10] B.J. Spencer, Physical Review B 59 (1999) 2011.
[11] K.H. Hansen, T. Worren, S. Stempel, E. Laegsgaard, M. Baumer, H.J. Freund, F. Besenbacher, I. Stensgaard, Physical Review Letters 83 (1999) 4120.
[12] W. Vervisch, C. Mottet, J. Goniakowski, European Physical Journal D 24 (2003) 311.
[13] A. Ramasubramanian, S.B. Shenoy, Journal of Applied Physics 95 (2004) 7813.
[14] H.C. Jeong, E.D. Williams, Surface Science Reports 34 (1999) 175.
[15] M. Giesen, Progress in Surface Science 68 (2001) 1.
[16] C. Herring, Physical Review 82 (1951) 87.
[17] A.F. Andreev, Zhurnal Eksperimentalnoi I Teoreticheskoi Fiziki 80 (1981) 2042.
[18] A.W. Searcy, Journal of Solid State Chemistry 48 (1983) 93.
[19] C. Jayaprakash, C. Rottman, W.F. Saam, Physical Review B 30 (1984) 6549.
[20] C. Rottman, M. Wortis, Physics Reports-Review Section of Physics Letters 103 (1984) 59.
[21] J.J. Metois, J.C. Heyraud, Surface Science 180 (1987) 647.
[22] M. Wortis, in: R.F.H. R. Vanselow (Ed.), Chemistry and Physics of Solid Surfaces VII, Berlin, 1988, pp. 367.
[23] M. Uwaha, P. Nozières, in: I. Sunagawa (Ed.), Morphology and Growth Unit of Crystals, Tokyo, 1989, pp. 17.
[24] E.D. Williams, N.C. Bartelt, Ultramicroscopy 31 (1989) 36.
[25] P. Nozieres, in: C. Godreche (Ed.), Solids far from Equilibrium, Cambridge, 1991, pp. 1.
[26] V.L. Pokrovsky, A.L. Talapov, Physical Review Letters 42 (1979) 65.
[27] C. Rottman, M. Wortis, J.C. Heyraud, J.J. Metois, Physical Review Letters 52 (1984) 1009.
[28] A. Pavlovska, D. Dobrev, E. Bauer, Surface Science 326 (1995) 101.
[29] J.M. Bermond, J.J. Metois, J.C. Heyraud, F. Floret, Surface Science 416 (1998) 430.
[30] J. Tersoff, E. Pehlke, Physical Review B 47 (1993) 4072.

[31]  D.J. Eaglesham, A.E. White, L.C. Feldman, N. Moriya, D.C. Jacobson, Physical Review Letters 70 (1993) 1643.
[32]  A. Emundts, H.P. Bonzel, P. Wynblatt, K. Thurmer, J. Reutt-Robey, E.D. Williams, Surface Science 481 (2001) 13.
[33]  M. Nowicki, C. Bombis, A. Emundts, H.P. Bonzel, P. Wynblatt, New Journal of Physics 4 (2002) 60.
[34]  S. Balibar, H. Alles, A.Y. Parshin, submitted to Reviews of Modern Physics (2005).
[35]  D.J. Eaglesham, F.C. Unterwald, D.C. Jacobson, Physical Review Letters 70 (1993) 966.
[36]  M. Giesen, C. Steimer, H. Ibach, Surface Science 471 (2001) 80.
[37]  S.V. Khare, S. Kodambaka, D.D. Johnson, I. Petrov, J.E. Greene, Surface Science 522 (2003) 75.
[38]  S. Kodambaka, S.V. Khare, V. Petrova, A. Vailionis, I. Petrov, J.E. Greene, Surface Science 523 (2003) 316.
[39]  R. Van Moere, H.J.W. Zandvliet, B. Poelsema, Physical Review B 67 (2003) 193407.
[40]  W.W. Mullins, G.S. Rohrer, Journal of the American Ceramic Society 83 (2000) 214.
[41]  G.S. Rohrer, C.L. Rohrer, W.W. Mullins, Journal of the American Ceramic Society 84 (2001) 2099.
[42]  N. Combe, P. Jensen, A. Pimpinelli, Physical Review Letters 85 (2000) 110-113.
[43]  J. Tersoff, A.W.D. Vandergon, R.M. Tromp, Physical Review Letters 70 (1993) 1143.
[44]  K. Thurmer, J.E. Reutt-Robey, E.D. Williams, M. Uwaha, A. Emundts, H.P. Bonzel, Physical Review Letters 8718 (2001) 186102.
[45]  K. Thurmer, J.E. Reutt-Robey, E.D. Williams, Surface Science 537 (2003) 123.
[46]  D.B. Dougherty, K. Thurmer, M. Degawa, W.G. Cullen, J.E. Reutt-Robey, E.D. Williams, Surface Science 554 (2004) 233.
[47]  P.L. Ferrari, M. Prahofer, H. Spohn, Physical Review E 69 (2004) 035102.
[48]  M. Degawa, E.D. Williams, in preparation (2005).
[49]  A. Pimpinelli, J. Villain, Physics of crystal growth, Cambridge ; New York, Cambridge University Press, 1998.
[50]  W.W. Mullins, Interface Science 9 (2001) 9.
[51]  P.J. Feibelman, Physical Review B 65 (2002) 129902.

**Figure Captions:**

Fig. 1 a) Illustration of a finite size crystallite with rounded regions between low index facets. A model for structure evolution is proposed in which the truncated shape, as illustrated by the dotted line, is treated as a constant volume crystal shape for purposes of modeling crystallite reshaping. Note here that the center of the crystallite (Wulff point) is in the interface plane of the modeled crystallite, a "free standing" boundary condition. b) Schematic drawing of a supported crystallite with a single flat facet bounded by rounded regions that terminate at the substrate. Here the Wulff point is not located in the interface plane and the height of the crystallite $z_h$ is determined by the adhesion energy.

Fig. 2 a) Dimensionless chemical potential $\tilde{\mu}_S$ as a function of dimensionless volume $\tilde{V}$ with different ratios of dimensionless adhesion energy $\tilde{E}_A =1.$, 1.5 and 2. at $\tilde{c}=0$. The parameter $\tilde{E}_A$ contains the temperature dependence through the temperature dependence of parameters $\beta$, $g$, $E_A$ and $\gamma_0$. As an example $\tilde{E}_A \approx 1.04$ for Pb at 27 °C. b) Dimensionless facet radius $\tilde{\rho}$ (or height $\tilde{z}_h$), contact slope $\tan\tilde{\varphi}_b$ and interface radius $\tilde{R}_b$ as a function of $\tilde{E}_A$ at constant volume at $\tilde{c}=0$. The dimensionless variables are given as eq. (A-1) and eq.(A-2)

Fig. 3 a) shows size scaled cross sections of the crystallites $\tilde{z}(\tilde{r})$ with same volume but different scaled $\tilde{c}$. The boundary condition is the "free standing" boundary condition ($E_A=\gamma_0$). The size scaling is given as eq. (A-3) and results in eq. (A-4). Note that these curzes are scaled by the facet radius and not by the crystallite volume itself. The maximum scaled $\tilde{c}$ shown is below the limiting value $\tilde{c} =1/6$.

Fig. 4 a) scaled crystallite height $\tilde{z}_h$, b) scaled interface radius $\tilde{R}_b$, c) contact slope $\tan\tilde{\varphi}_b$ d) scaled crystallite volume and e) scaled excess chemical potential $\tilde{\mu}_S$ are shown as a function of scaled $\tilde{c}$ with constant volume.  Size and adhesion energy (temperature) scaling is given as eq. (A-3) and (A-6) and results in eq. (A-5).  The precise values of the universal scaled values at c=0 are given in Table 1.

Fig. 5 gives the crystallite free energy as a function of $c$ up to $c_{max}$ with same volume for a) the "free standing boundary" condition ($E_A=\gamma_0$) Thermodynamic parameters $\gamma_0$=1.7 eV/nm$^2$[51], $\beta$=0.34 eV/nm and $g$=0.65 eV/nm$^2$ [32,33] where used for Pb at 27 ˚C.  b) with adhesion energy $E_A$=0.5$\gamma_0$ (open) and c) with $E_A$=1.5$\gamma_0$ (open), respectively.  The surface free energy alone (filled) is also given as closed symbols in b) and c).  d) gives the scaled total surface free energy $\tilde{F}$ as a function of $\tilde{c}$ up to 1/6.  Size and adhesion energy scaling is given as eq. (A-6)

Fig. 6  A log-log plot of a 3 parameter fits to the PT-shape eq. (10) and the critical state shape eq. (16) to the measured cross section of the defect-free Pb crystallite, taken with an STM at ~ 300K.  The fits yield ratios of $g/\beta$=17.2 nm$^{-1}$ and $g/\beta$=3.6 nm$^{-1}$, respectively.

Fig. 7 scaled barrier height $\Delta\tilde{E}$ as a function of $\tilde{c}$ with constant volume and different adhesion energies ($E_A$=0.5$\gamma_0$, $\gamma_0$, 1.5$\gamma_0$) using the method by Rohrer and Mullins [40,41]. Size scaling is given by eq. (A-3) and results in eq. (A-5)

Table 1. Table of values calculated from the analytical solutions (eqs. (A-5)) for different $\tilde{c}$, also given graphically in Fig. 4 and 7.

| $\tilde{c}$ | $\tilde{z}_h$ | $\tan \tilde{\varphi}_b$ | $\tilde{\mu}_S$ | $\Delta \tilde{E}$ |
|---|---|---|---|---|
| -0.08 | 0.806 | 0.08 | 2.48 | 2.77 |
| -0.04 | 0.893 | 0.04 | 2.24 | 2.16 |
| 0 | 1 | 0 | 2 | 1.57 |
| 0.04 | 1.14 | -0.04 | 1.76 | 1.03 |
| 0.08 | 1.32 | -0.08 | 1.52 | 0.559 |
| 0.12 | 1.56 | -0.12 | 1.28 | 0.192 |
| 0.16 | 1.92 | -0.16 | 1.04 | 0.00483 |

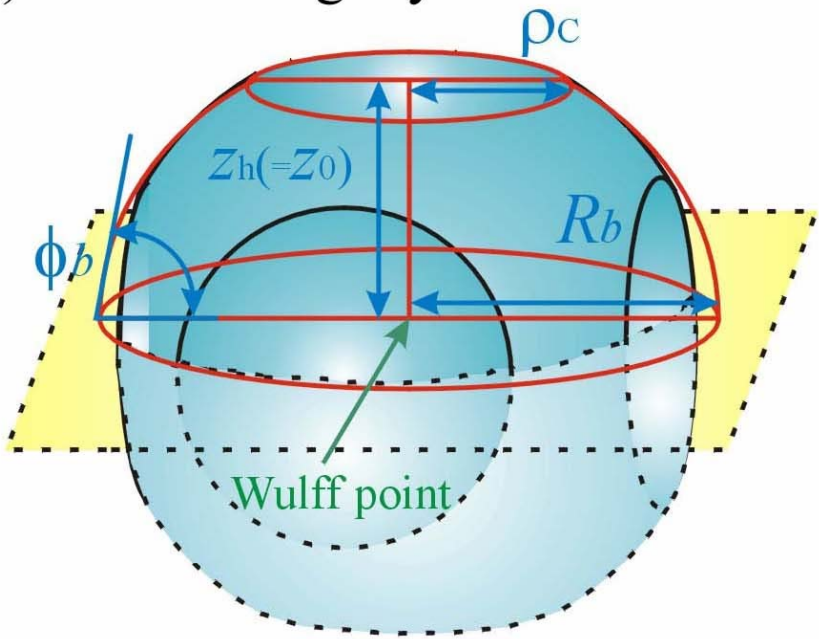

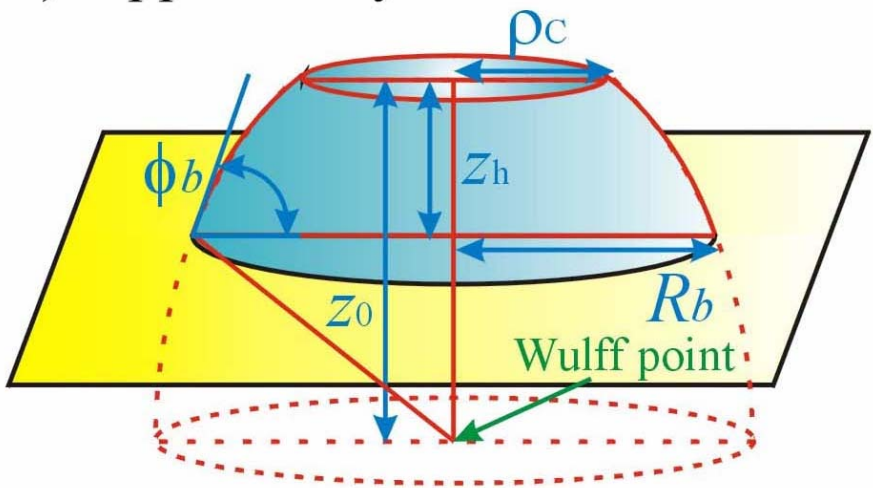

Fig. 1 a) Illustration of a finite size crystallite with rounded regions between low index facets. A model for structure evolution is proposed in which the truncated shape, as illustrated by the dotted line, is treated as a constant volume crystal shape for purposes of modeling crystallite reshaping. Note here that the center of the crystallite (Wulff point) is in the interface plane of the modeled crystallite, a "free standing" boundary condition. b) Schematic drawing of a supported crystallite with a single flat facet bounded by rounded regions that terminate at the substrate. Here the Wulff point is not located in the interface plane and the height of the crystallite $z_h$ is determined by the adhesion energy.

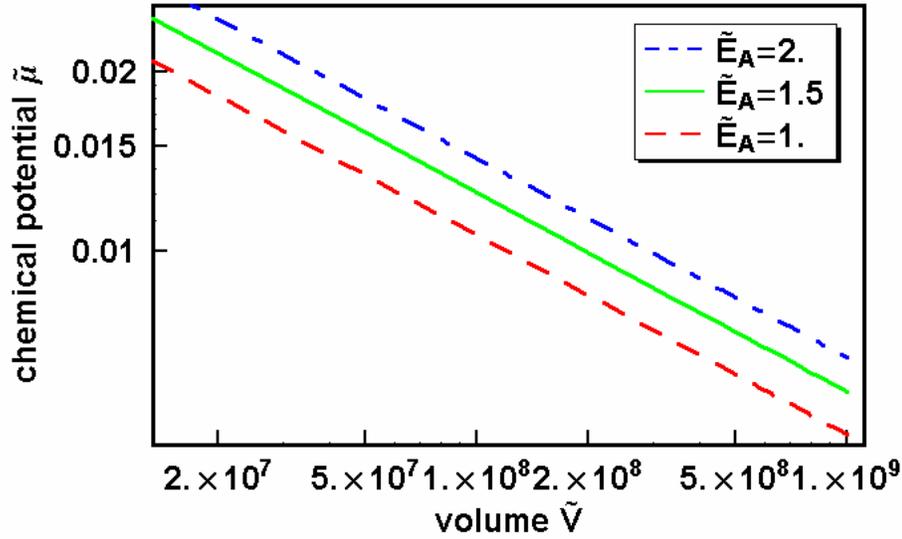

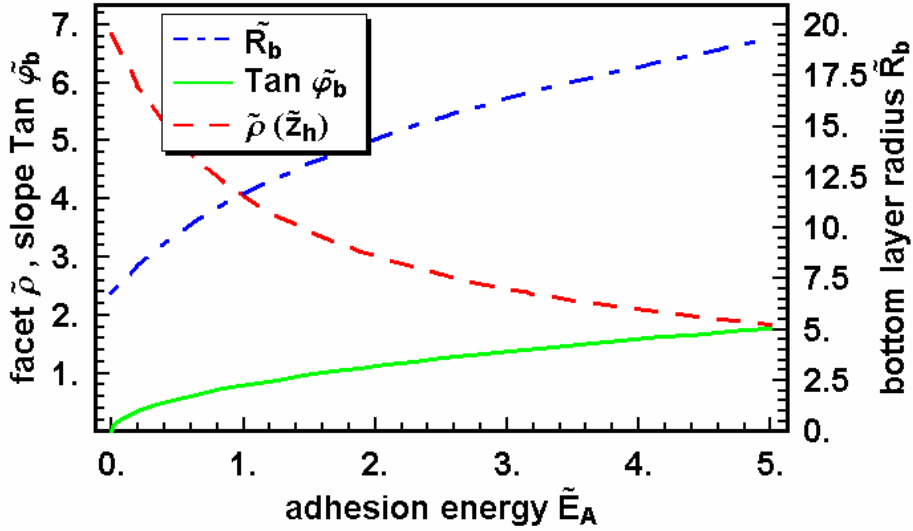

Fig. 2 a) Dimensionless chemical potential $\tilde{\mu}_S$ as a function of dimensionless volume $\tilde{V}$ with different ratios of dimensionless adhesion energy $\tilde{E}_A$ =1., 1.5 and 2. at $\tilde{c}$=0. The parameter $\tilde{E}_A$ contains the temperature dependence through the temperature dependence of parameters $\beta$, $g$, $E_A$ and $\gamma_0$. As an example $\tilde{E}_A \approx 1.04$ for Pb at 27 °C. b) Dimensionless facet radius $\tilde{\rho}$ (or height $\tilde{z}_h$), contact slope $\tan\tilde{\varphi}_b$ and interface radius $\tilde{R}_b$ as a function of $\tilde{E}_A$ at constant volume at $\tilde{c}$=0. The dimensionless variables are given as eq. (A-1) and eq.(A-2)

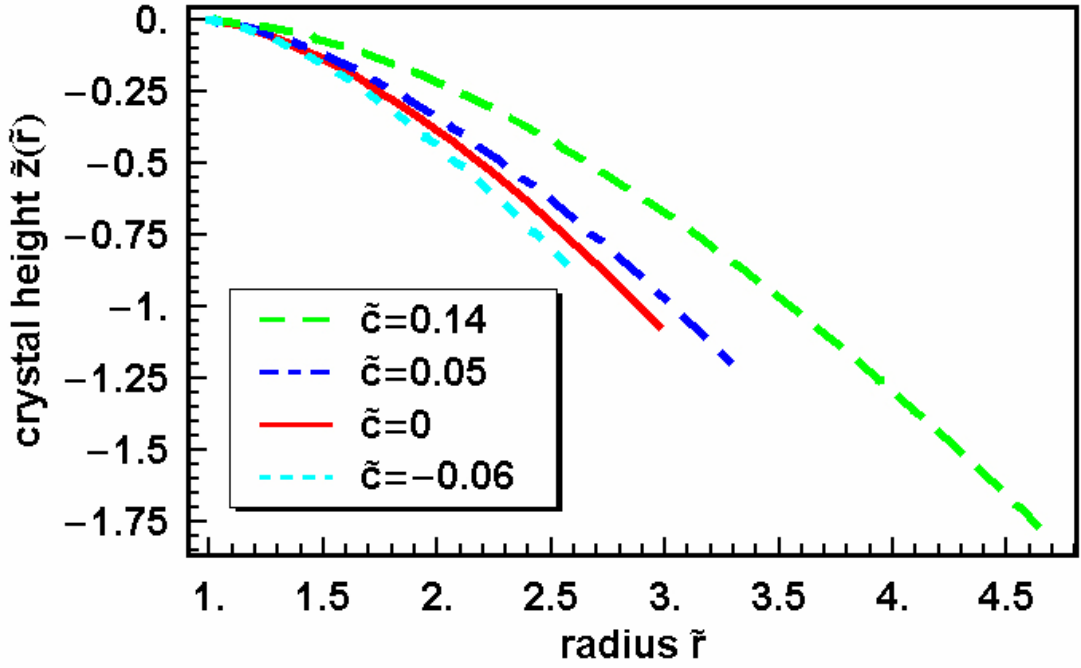

Fig. 3 a) shows size scaled cross sections of the crystallites $\tilde{z}(\tilde{r})$ with same volume but different scaled $\tilde{c}$. The boundary condition is the "free standing" boundary condition ($E_A=\gamma_0$). The size scaling is given as eq. (A-3) and results in eq. (A-4). Note that these curzes are scaled by the facet radius and not by the crystallite volume itself. The maximum scaled $\tilde{c}$ shown is below the limiting value $\tilde{c}=1/6$.

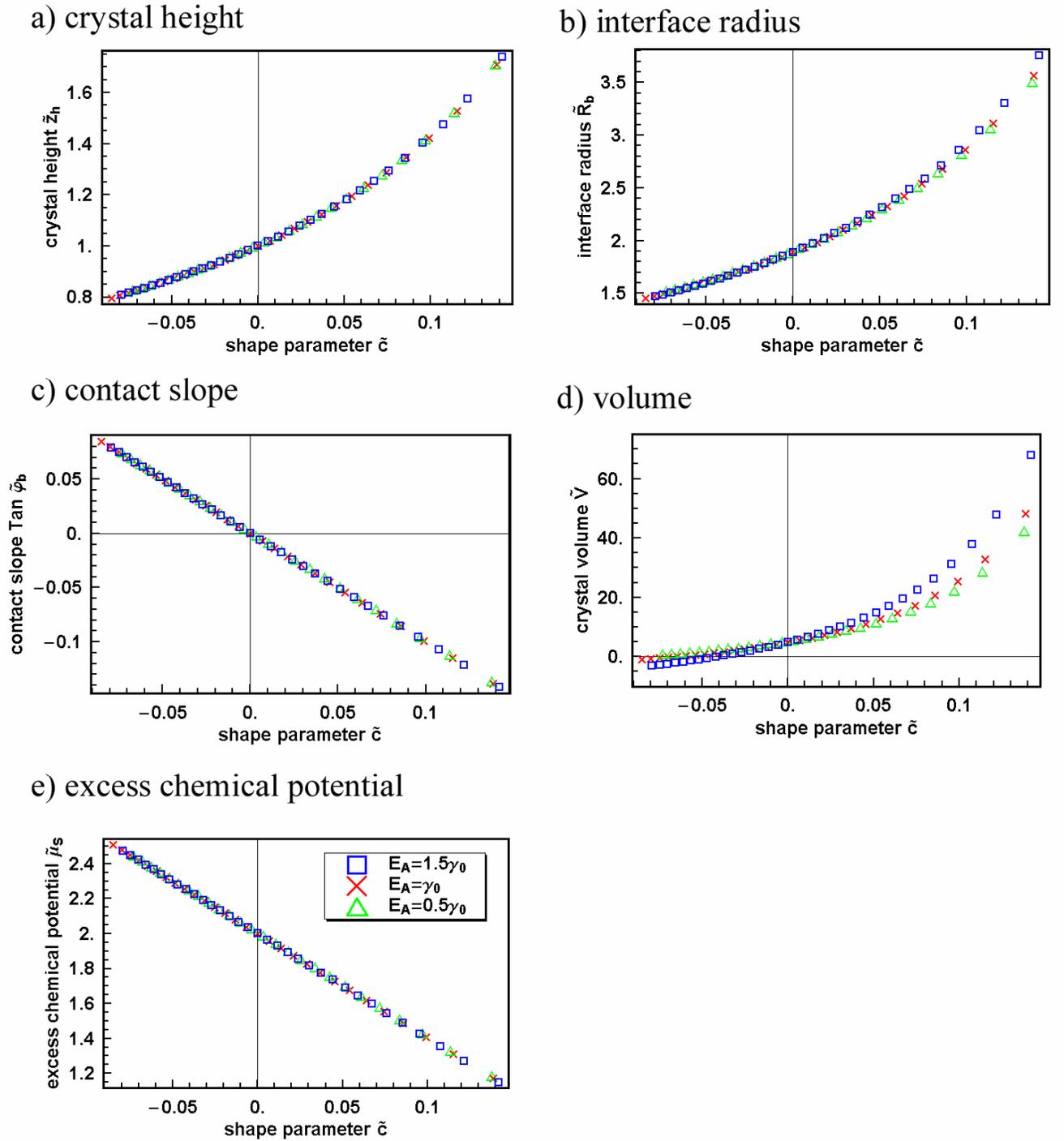

Fig. 4 a) scaled crystallite height $\tilde{z}_h$, b) scaled interface radius $\tilde{R}_b$, c) contact slope $\tan\tilde{\varphi}_b$ d) scaled crystallite volume and e) scaled excess chemical potential $\tilde{\mu}_S$ are shown as a function of scaled $\tilde{c}$ with constant volume. Size and adhesion energy (temperature) scaling is given as eq. (A-3) and (A-6) and results in eq. (A-5). The precise values of the universal scaled values at c=0 are given in Table 1.

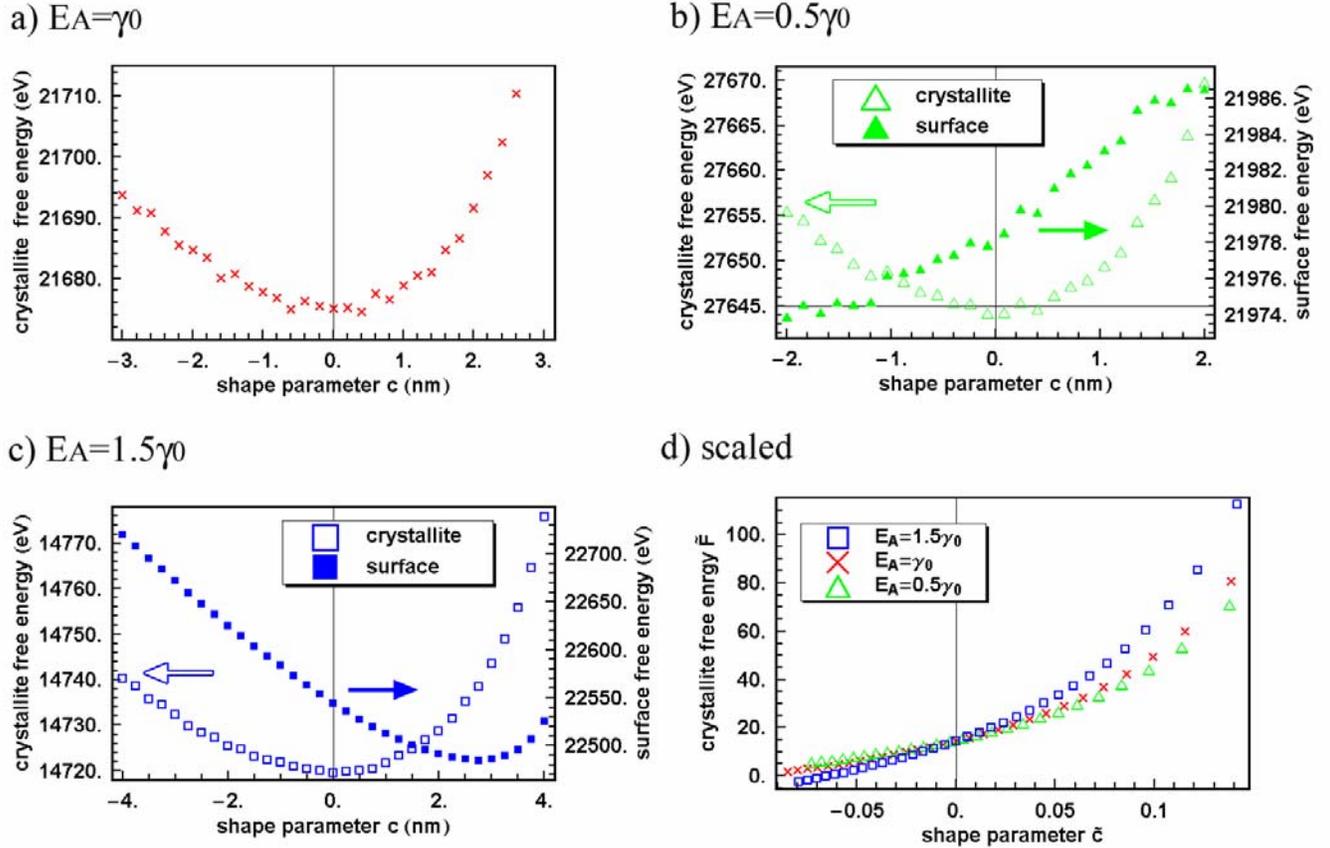

Fig. 5 gives the crystallite free energy as a function of $c$ up to $c_{max}$ with same volume for a) the "free standing boundary" condition ($E_A=\gamma_0$) Thermodynamic parameters $\gamma_0=1.7$ eV/nm$^2$[51], $\beta=0.34$ eV/nm and $g=0.65$ eV/nm$^2$ [32,33] where used for Pb at 27 ˚C. b) with adhesion energy $E_A=0.5\gamma_0$ (open) and c) with $E_A=1.5\gamma_0$ (open), respectively. The surface free energy alone (filled) is also given as closed symbols in b) and c). d) gives the scaled total surface free energy $\tilde{F}$ as a function of $\tilde{c}$ up to 1/6. Size and adhesion energy scaling is given as eq. (A-6)

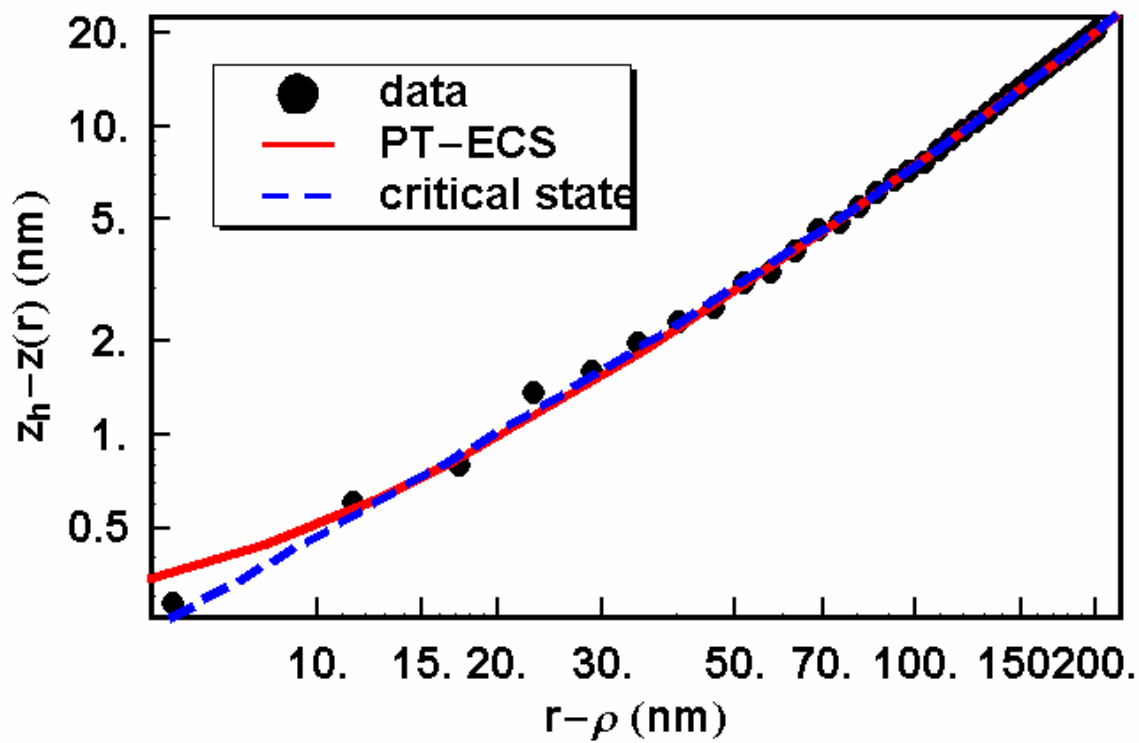

Fig. 6 A log-log plot of a 3 parameter fits to the PT-shape eq. (10) and the critical state shape eq. (16) to the measured cross section of the defect-free Pb crystallite, taken with an STM at ~ 300K. The fits yield ratios of $g/\beta$=17.2 nm$^{-1}$ and $g/\beta$=3.6 nm$^{-1}$, respectively.

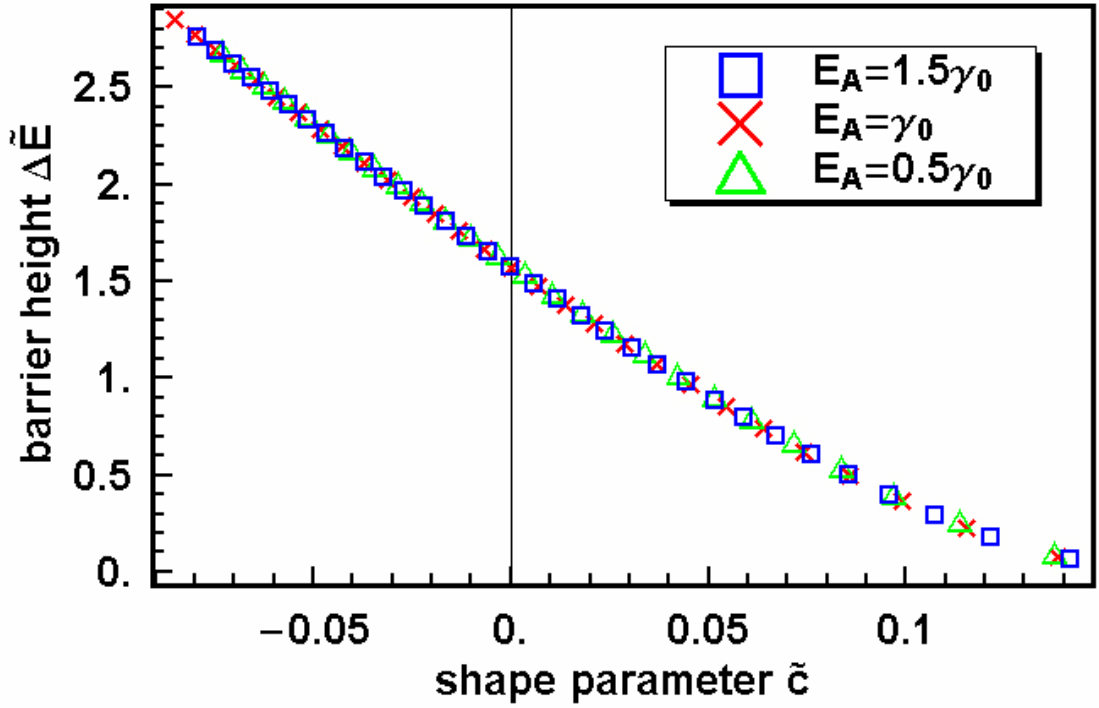

Fig. 7 scaled barrier height $\Delta \tilde{E}$ as a function of $\tilde{c}$ with constant volume and different adhesion energies ($E_A$=0.5$\gamma_0$, $\gamma_0$, 1.5$\gamma_0$) using the method by Rohrer and Mullins [40,41]. Size scaling is given by eq. (A-3) and results in eq. (A-5)